\documentclass[a4paper,aps,prl,showpacs,twocolumn,superscriptaddress]{revtex4}

\usepackage{graphicx,t1enc}

\begin{document}

\date{\today}

\title{A model for the emergence of geopolitical division}

\author{M. N. Kuperman}

\affiliation{Centro At{\'o}mico Bariloche and Instituto Balseiro, 8400 S. C. de Bariloche, Argentina\\
Consejo Nacional para las Investigaciones Cient\'{\i}ficas y
T\'ecnicas, Argentina}

\begin{abstract}
In this work, we present a model based on a competitive dynamics
that intends to imitate the processes leading to some
characteristics of the geopolitical division. The model departs
from very simple principles of geopolitical theory and geometrical
considerations, but succeeds to explain general features related
to the actual process. At the same time, we will propose an
evolutionary explanation to the fact that most capitals (in
Eurasia) are located far from the borders or coasts and, in many
cases, close to the barycenter of the respective countries.
\end{abstract}

\maketitle

\section{Introduction}

The combination of both a suitable theory and empirical data can
render macro historical predictions possible. This is one of the
statement enounced by R. Collins in \cite{col2}, where he accounts
for a successful prediction of the breakup of the Soviet Union.
The suitable theory that frames his model \cite{col1} is a
geopolitical theory for the power of the states. The history of
geopolitical theory started  in Germany at the beginning of the
20th century with the works by Weber \cite{web} on the development
of the state. But it was not until the 70's, after many years of
declining interest on the subject, that the studies on this area
of knowledge flourished \cite{hep}. Among all the processes
studied by geopolitical history, the formation of the modern
states system emerges as one of the most relevant, involving
feedback processes between the economy, the society and the
politics.

The geopolitical configuration of the continents and its evolution
was a determinant factor on the economic success of several areas
of the world. At the same time it was critical in sealing the
future political regimes of the nations. As complex as the spatial
evolution of states might be, Collins \cite{col1} proposed a
simple theoretical model to deal with the main aspects involved in
the process of expansion and contraction of the territorial power
of sates. The model is based on five principles that will be
summarized in the following paragraph. These principles describe
the factors that promote the expansion or the collapse of a
nation. On one side, the expansion of states is favored by the
geographical size and availability of  resources and by
geopositional advantages, e.g.  countries with fewer enemies
expand at the expenses of other countries with more enemies on the
borders. However, on the other side, a strain of resources may
arise as a result of overextension, leading to the disintegration
of the state. In the same model Collins assumes that states in the
middle of a geographical region tend to fragment into smaller
unities over time. Finally, the model affirms that some cumulative
processes give place to long term simplification, with massive
wars between a few contenders. We can give an example of the
processes foreseen by  these principles. During historical periods
previous to the technological explosion derived from the
industrial revolution the states increased their power through
conquest and expansion. But this process entailed an increment in
the costs of administration and military defense. Therefore,
whereas the seek for power promoted the expansion, the increment
in size and in administrative duties imposed critical constrains.
At the initial stages of expansion the increments in the wealth of
a successful state exceeded the incremental costs of defending the
additional new territory. However, when borders reached a critical
distance from the center of power those costs rose faster than the
benefits. As  the survival of a polity depended on its military
power to neutralize the constant threat posed by contiguous
states, the protection of the borders was crucial \cite{nor}.

In \cite{jon} the author analyzes the interplay between the
evolution of the geopolitical structure of Europe towards a system
of competing states and the social development. This competing
structure promoted the creation of a pluralistic environment that
succeeded to constrain the power of the ruling classes, as well as
the power of the states on the individuals, helping to promote
institutions and legal systems that ultimately provided greater
freedom than did monolithic empires. Jones affirms that the
ultimate configuration of the state system in Europe can not
respond to mere geometric aspects and funds his affirmation on the
analysis of a collection of complex and interconnected facts that
should have influenced on the whole geopolitical process. These
conclusions seem to contradict the ideas of Collins, whose model
is mainly based on geometrical considerations. The goal of this
work is to show that although geometry can not account for the
whole geopolitical process, it does impose some constrains with
important effects on the geopolitical division process. When the
geometry of a given polity is closely related to other indicators
such as the military power and the availability of resources some
basic aspects of the process of political division can be derived
from a model solely based on topological principles.

This task was partially accounted by Artzrouni and  Komlos
\cite{art}, where they analyzed the spatial evolution of the
European state system throughout a period extending roughly from
A.D. 500 to A.D. 1800. Following this previous work, we generalize
the model to include some additional aspects, such as the location
of the center of power controlling each of the states and the
analysis not only the final configuration but of the temporal
evolution towards the steady state.

\section{The Model}
The model presented here is associated to a previous work  by
Artzrouni and  Komlos \cite{art}, where the authors introduced a
spatial predatory model to trace the evolution of the political
borders in Europe, from  500 A.D. to 1800 A.D.  The dynamics of
the model is based on the fact that any political unit has a
natural tendency to expand by conquering neighboring areas. A
first simplistic assumption is that its perspective of success or
failure in combat will depend on its  military power. Following
the principles stated in \cite{col1} the present model considers
that the military power increases with the size of a state but
decreases as a result of the over-extension of the borderline and
the increment of the distance between the center of power and the
borders. Though the present model represents an oversimplification
of the real historical process still captures the most essential
features. Among the neglected features we can mention the
evolution of military technology, as well as the inhomogeneities
in the distribution of the population and other economic
indicators. In this version of the model we will consider the
expansion or collapse by annexation but we will neglect the
segregation of big polities into smaller units. Each country
struggles for its survival by combating its neighbouring rivals.
During the initial stages of the evolution this process leads the
successful countries to expand and the defeated states to shrink
and disappear. The gains in wealth and power associated to the
expansion exceed the incremental costs of defending the annexed
territory. Some defeated countries eventually disappear as a
continuous loss of territories undermines their military power and
can not withstand further attacks from their opponents. At a later
stage the length of the expanding borders increase as well as
their distances to the center increase, making the costs to
protect them go beyond the benefits of expansion. At this point,
the states face the dilemma of maintaining or stopping  an
expansive process involving costs that become increasingly
difficult to overcome.

Based on the previous considerations, Artzrouni and  Komlos
\cite{art} define the military power of a state, $P$, as a
quantity depending on the geometry of each state. It is expressed
as a function of two variables.: the area $A$ and perimeter $F$ of
a country. The states are located on top of a square lattice. The
area $A$ is simply the number of nodes of the grid occupied by the
country, whereas the perimeter is the number of bordering nodes.
In turn, $P(A,M)$ is defined as
\begin{equation}
P(A,M)= \frac{A}{\delta + \exp( \gamma F + \beta )},
\label{power1}\end{equation}
 where $\delta$, $\beta$, and
$\gamma$ are positive parameters.

One of the modifications introduced in the present work is the
addition of fixed centers of power or capitals with a crucial role
on  the dynamics of the system. The military power $P$ will depend
on the area $A$ and on a quantity $M$ that measures the distance
between the frontier points and the center of power. Physically,
this quantity is associated to a power of the moment of inertia of
the border of a country, relative to its capital.
\begin{equation}M_k=\sum_{b\in\Omega_k} [(i_b-i_{c_k})^2+(j_b-j_{c_k})^2]^\alpha,
\label{inert}\end{equation} where $\Omega_k$ is the border of the
country $k$,  $i_b$ and $j_b$ are the coordinates of a border node
$b$  and $i_{c_k}$ and $j_{c_k}$ the coordinates of the capital.
In the previous model two different geomorphological aspects have
been  considered: The first one is the existence of natural
barriers due to topological aspects of the terrain, that only
inhibit the interaction between adjacent nodes. The second one is
the existence of coasts, generally more easy to defend. We will
account only for the effect of the coastal borders on $M$ through
a weighting  factor $\kappa<1$. Thus, Eq.(\ref{inert}) results in
\begin{eqnarray}M_k&=&\sum_{b\in\Omega^1_{k}} [(i_b-i_{c_k})^2+(j_b-j_{c_k})^2]^\alpha+ \nonumber\\
&&\sum_{b\in\Omega^2_{k}}
\{\kappa[(i_b-i_{c_k})^2+(j_b-j_{c_k})^2]\}^\alpha,
\label{inert2}\end{eqnarray} where $\Omega^1$ takes into account
the inland border and $\Omega^2$ the coastal border nodes.
$P(A,M)$ is defined as
\begin{equation}
P(A,M)= \frac{A}{\exp( \gamma M )}. \label{power2}
\end{equation}
According to the previous discussion, we want the  military power
to decide the outcome of a combat, though we are interested  in
preserving  a certain degree of stochasticity as well. Therefore,
we follow the prescription proposed in \cite{art}. After measuring
the $P$ values of the two confronting countries we call $P_h$ and
$P_l$ the higher and lower $P$ values respectively. This
identification is irrelevant in case of equality and can be
randomly done as shown later. Next we assign the victory to the
country with the higher military power with a probability
$$\phi_h = 1 - 0.5 \exp (-k (P_h/P_l - 1)),$$ where $k$ is
the parameter that tunes the deterministic character of the
dynamics. The higher $k$ is the higher the possibility of the
strongest country to win will be. The weakest country wins with a
probability $\phi_l=1-\phi_h$. In case of equality the
probabilities are $\phi_l=\phi_h=0.5$. The same is true when
$k=0$, and the systems evolves in a random way.

The analysis of  Eqs. (\ref{power1}) and (\ref{power2}) clarifies
the idea that $P$ is a quantity that contains information about
the military power of a state and about how efficiently this power
can be used. The area $A$ favors the economic power and population
size.  But beyond a certain threshold level of spatial expansion
the military power becomes less effective  since more resources
are needed for the defense of the enlarged frontier. On one hand
$P$ is a monotonic increasing function of $A$, indicating that a
greater area represents a potentially higher power. On the other
hand, a bigger area can inhibit the efficient exertion of that
power by effects of the state´s geometry. This fact is taken into
account in the dependence of $P$ on $F$ or $M$. As an example, we
will consider a state represented by  $n^2$ territorial unities.
In two extreme situations the state can be a $n\times n$ square or
a $n^2\times 1$ rectangle. Though both state have the same area,
the rate perimeter - surface is lower for the square state,
consequently, according to Eq. ((\ref{power1})) its military power
is higher. A similar conclusion can be derived when considering
Eq. (\ref{power2}). According to Steiner theorem, the moment of
inertia of a plane figure is minimum when it is calculated about
an axis located at the center of mass. Therefore, among all the
states with  identical shape, the one with its capital being the
closest to the center of mass will present the lowest value of
$M$. A possible observation to the situation described above is
the fact that borders closer to the center of power should be
easier to protect that those far away. At the same time, a wiser
military strategy will tend to focus the defense of the country on
the borders but without neglecting the capital. To consider this
facts we introduce a further modification of Eq. (\ref{power2}),
by including a local term that accounts for the relative distance
between the capital and the border points involved in a specific
conflict, i.e. the points at the border with the eventual opponent
country. The military power presents then a global term plus a
local factor associated with the conflictive border. We redefine
$P$ as
\begin{equation}
P(A,M)= \frac{A}{\exp( \gamma MM_l )}.
\label{power3}
\end{equation}
The calculation of  $M_l$ is analogous to that of  $M$ but restricted to the border under conflict, i.e.
$\Omega^1$ and  $\Omega^2$ in Eq.(\ref{inert2}) refer only to the partial border.

As we will show later, the dynamics of this system can be
associated to a cluster growth process. Although there is a vast
literature on this area \cite{lan,hee,vic,sta,her,van} and some of
these works might show some common features with the one described
in this work,  the present model includes some unique features
closely related to the particular problem here analyzed. In the
following section we discuss the results obtained in each case

\section{Results}
\begin{figure}[!h] \centering \resizebox{9cm}{!}
{\rotatebox[origin=c]{0}{\includegraphics{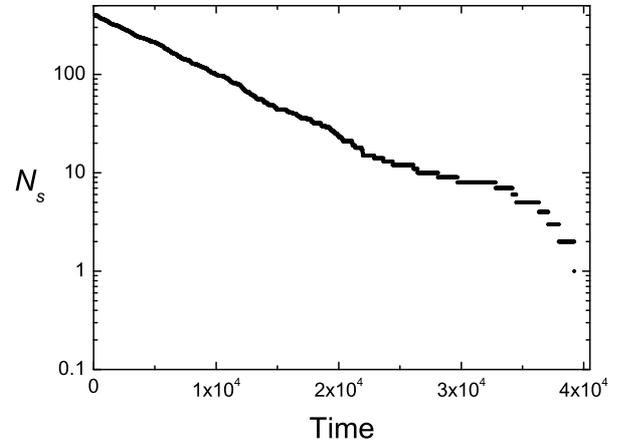}}}
\caption{$N_s$ vs. time for a random dynamics.} \label{rand}
\end{figure}

Throughout all the simulations we have maintained the same initial
condition. At the beginning the system is composed by 400 states,
each one being a square comprising  $3 \times 3$ surface unities
with the capital located at the barycenter. The whole system is a
square lattice of $60 \times 60$ surface unities. During the first
stages of the simulation, the dynamics of the system  is purely
random, until some differences among the shapes of the states
arise,  playing a defining role in the further evolution of the
system. In order to get a correct interpretation of our results,
we analyze first the behaviour of a system  driven by a random
dynamics throughout the whole simulation. This can be very easily
achieved by considering that the  outcome of a fight is
uncorrelated to the military power of the confronting countries.
In this case the winner is randomly chosen. Figure \ref{rand}
summarizes the main findings: the dynamics of the model drives the
system to the conformation of a unique big country having
conquered all the available territory. The collapse of smallest
countries produces an exponential decay in the number of survival
countries $N_s$, with a slowing down when the number of remaining
countries is around 10. At this point the disappearance of
countries takes much longer. Finally, the system converges to its
only steady state.

This trivial stochastic evolution turns much more interesting
when, as proposed in the model,  it is the value $P$ what defines
the victorious country. According to what was discussed before, we
expect the states displaying values of  $P$ monotonically related
to their sizes at the beginning. After a transient letting to the
disappearing of the majority of the countries and the conformation
of a bunch of powerful sates, the effect of the critical size must
be apparent. At this point the increasing value of $F$ or $M$
dominates the evolution of the states, making their $P$ values to
decrease in case of overextension.
\begin{figure}[!h] \centering \resizebox{9cm}{!}
{\rotatebox[origin=c]{0}{\includegraphics{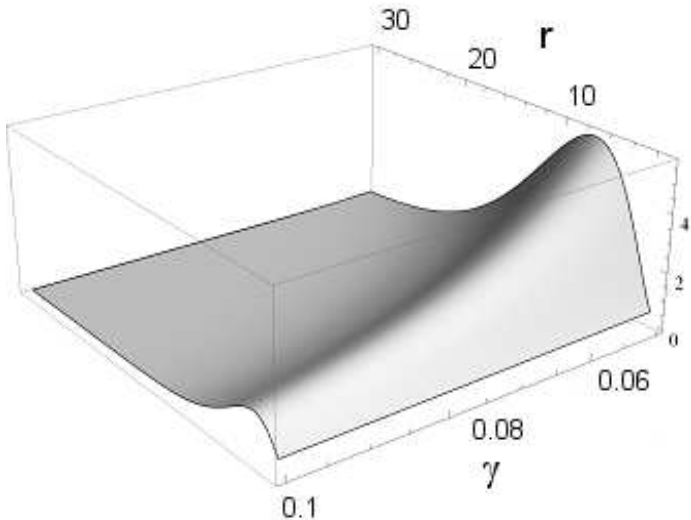}}}
\caption{$P(r,\gamma)$ for the case associated to
Eq.(\ref{power1}), $\alpha=\beta=1$}. \label{pb1}
\end{figure}
The previous affirmations can be supported by analytical results.
Considering Eqs. (\ref{power1}) and (\ref{power2}), we can state
that the optimal shape adopted by any state is the circular one.
In our analysis, and only for simplicity,  we will consider the
octagonal shape. Furthermore, if the location of the capital is
taken into account, $P$ is maximized when the former is located on
the center of mass of the corresponding state. We define $r$ as
the radius of the circle circumscribing the octagon. As discussed
before, an expanding state will increase its military power until
a critical value of $r$ is attained. At this point the effects of
the overextension will start to be noticeable. We can estimate the
maximum attainable value of $P$ at this critical radius $r_c$.

\begin{figure}[!h] \centering \resizebox{9cm}{!}
{\rotatebox[origin=c]{0}{\includegraphics{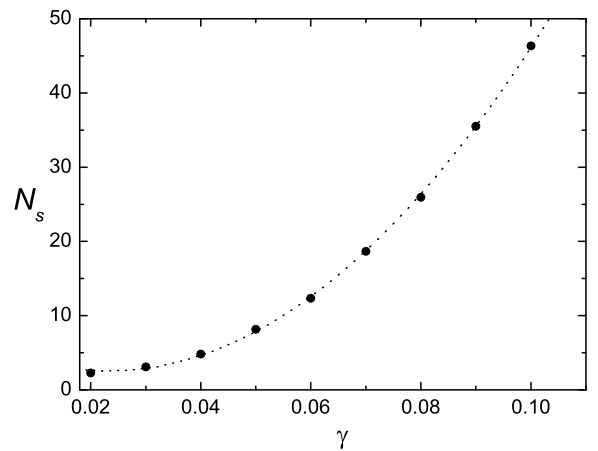}}}
\caption{$N_s$ vs. $\gamma$ for the case associated to
Eq.(\ref{power1}), $\alpha=\beta=1$. The dashed line corresponds
to fitting the data with a parabola}. \label{nvsg1}
\end{figure}

\begin{figure}[!h] \centering \resizebox{9cm}{!}
{\rotatebox[origin=c]{0}{\includegraphics{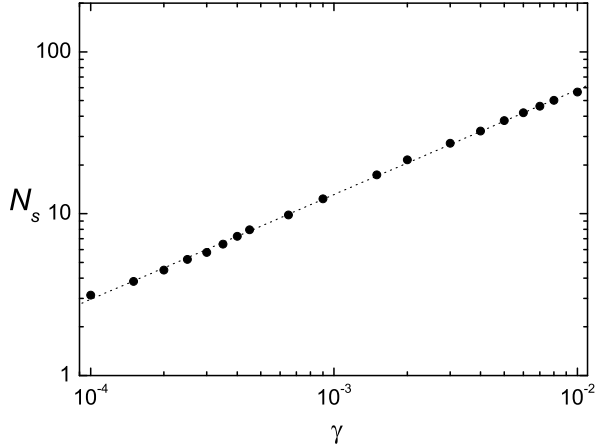}}}
\caption{$N_s$ vs. $\gamma$ for the case associated to
Eq.(\ref{power2}), $\alpha=1$. The scales are logarithmic. The
dotted line has a slope equal to 2/3}. \label{nvsg2}
\end{figure}

If we analyze the Eq.(\ref{power1}) considering that a state
expands preserving an octogonal shape and a centered capital, we
can find an approximate value of $r_c$, the point when the maximum
military power is achieved. We note that the quantities $A$, $F$
and  $P$ can be written now in terms of $r$.
\begin{eqnarray} A(r)& = &2 (2\sin(\pi/8) r)^2 (1
+\sqrt{2})\nonumber \\
F(r)&=&16 \sin(\pi/8) r  \label{pbr1} \\
P(r)&=&\frac{A(r)}{(\alpha + \exp(\gamma P(r) + \beta)} \nonumber
\end{eqnarray}
Figure \ref{pb1} shows the value of $P(r)$ according to the
expression in Eq. (\ref{pbr1}) in the $(r,\gamma)$ plane . The
existence of critical radius $r_c$ is evident.
Deriving the last of Eqs. (\ref{pbr1}) with respect to $r$ let us
find $r_c$, getting the following expression
\begin{equation}
\mbox{e}^{\beta + 16 r_c \gamma \sin(\pi/8)} ( 8 r_c \gamma
\sin(\pi/8)-1)-\alpha=0
\end{equation}
with a solution for $r_c$ in terms of Lambert's  $W$ function,
\begin{equation}
r_c=\frac{2 + W [2\alpha \exp(-2-\beta)]}{16\sin(\pi/8)\gamma}
\label{rmax1}
\end{equation}
The former expression shows us that the critical radius is
inversely proportional to $\gamma$. If we consider that the mean
size of the countries is inversely proportional to the square of
the radius, we get the following relation: $N_s \propto \gamma^2$.
This is precisely what we get when plotting the numerical data on
the $(N_s,\gamma)$ plane and we fit them with a parabola, as shown
in Fig. \ref{nvsg1}.
When we take the location of the capital into account we get the
following expressions:
\begin{eqnarray} A(r)& = &2 (2\sin(\pi/8) r)^2 (1
+\sqrt{2})\nonumber \\
M(r)&=&2(2+\sqrt{2}) \sin(\pi/8) r^{ 2 \alpha + 1}\\
B(r)&=& \frac{A(r)}{\exp(\gamma M(r)) } \nonumber \label{pbr2}
\end{eqnarray}
In this case the critical radius is inversely proportional to
$\gamma^{1/(2 \alpha + 1)}$, therefore $N_s \propto
\gamma^{\frac{2}{2 \alpha + 1}}$. Fig. \ref{nvsg2} shows an
example of the fitting of numerical data when $\alpha=1$. The
fitting shows us that $N_s \propto \gamma^{\frac{2}{3}}$

This  analysis is not applicable to Eq. (\ref{power3}) because it
involves not only the geometry of the analyzed state, but also
that of its neighbours as well as local considerations.

All the data on Figs. \ref{pbr1} and \ref{pbr2} correspond to
averages over 1000 realizations for each point. The number of
surviving states $N_s$ reaches a stationary value, maintained over
at least 50000 time steps, ten times longer that the time it takes
the system to reach the steady value. Although the number of
surviving states remains constant, the system is not frozen. We
can observe to changes in the shapes and sizes of these surviving
countries due to persisting combats. Nevertheless, these fights do
not change significatively the conformed map. These results let us
confirm that the system evolves, with some countries expanding at
cost of the collapse of others, until a metastable situation is
achieved.

At his point the analysis of the model described by Eq.
(\ref{power1}) is almost through except for the study of the shape
and size distribution of the surviving countries.
\begin{figure}[!h] \centering \resizebox{9cm}{!}
{\rotatebox[origin=c]{0}{\includegraphics{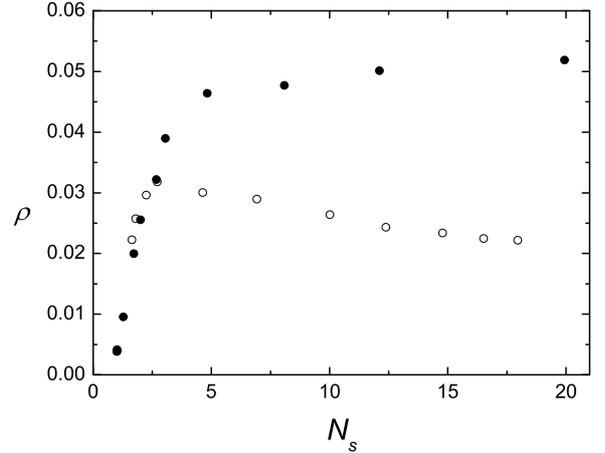}}}
\caption{$\rho$ vs. $N_s$, the number of surviving countries.
Filled dots: Eq.(\ref{power1}), Empty dots: Eq.(\ref{power2}).
$\alpha=\beta=1$. } \label{simet}
\end{figure}
To statistically analyze the geometry of the surviving countries
we measure how far from a circle or octagon their shapes are. We
consider a quantity $\rho$ herewith defined. First we measure the
distance $d_{i,0}$ between each point $i$ on the borderline of the
country and its center of mass. Then we calculate $\rho$ as the
ratio between the variance of this distances and the area of the
country, $A$. If a country is circular, this quantity is equal to
cero, while it is $\approx 0.00112$ for an octagon. The expression
for $\rho$ is then
\begin{equation}
\rho=\frac{1}{A} \left( \frac{1}{N_\Omega} \sum_{i \in \Omega}
(d_{i,0})^2 - (\frac{1}{N_\Omega} \sum_{i \in \Omega}d_{i,0})^2
\right)
\label{ro}
\end{equation}

In Fig. \ref{simet} we plot $\rho$ versus the number of surviving
states, $N_s$. The filled dots correspond to the present case. We
observe the sharp transition as the number of surviving countries
grows from 1 to 3 . The lower value, $\rho \approx 4\times 10^3$
corresponds to a unique squared country limited by the shape and
size of the lattice. The value of $\rho$ then stabilizes around
$5\times 10^2$. As a reference,  a rectangular, almost
1-dimensional country of the same  size when $\gamma=0.1$ would
have $\rho \approx 10$.  The variation in size, whose mean value
is associated to the  mean number of surviving states depicted in
Fig. \ref{nvsg1},  is around 50\% when the number of surviving
countries is over 5, then it decreases abruptly to zero. The
introduction of  fixed capitals let us add to the previous
analysis the study of their locations and of the centrality of the
respective countries. One of the aspects we wanted to analyze by
means of the modification introduced to the original model was
precisely where the capitals of the surviving countries are
located. Figure \ref{simet} also shows the behaviour of $\rho$ in
this case, plotted with empty dots. The most evident fact is that
the values are lower than in the previous case, indicating that
the countries adopt a more symmetric and circular shape. The
results do not differ too much if we consider either Eq.
(\ref{power2}) or Eq. (\ref{power3}).

When considering the location of the capital, not only we can
measure $\rho$ but we can  calculate the distance between the
center of mass and the capital. we choose the mean radius of the
country to compare the previous distance with a characteristic
length. We define $\delta_{c0}=d_{c,0}/r_m$, where $d_{c,0}$ is
the distance between the capital and the center of mass and $r_m$
is the mean radius.
\begin{figure}[!h] \centering \resizebox{9cm}{!}
{\rotatebox[origin=c]{0}{\includegraphics{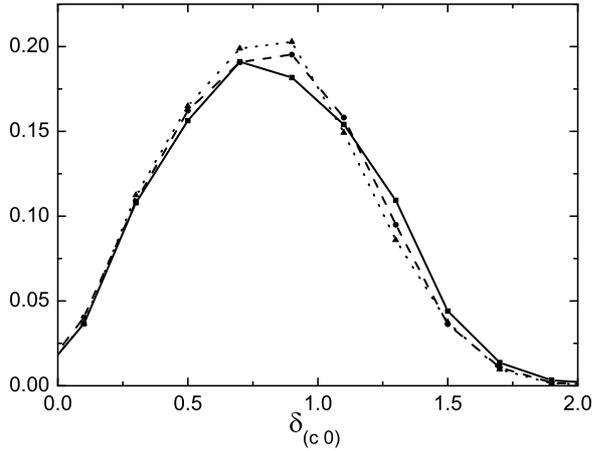}}}
\caption{ Distribution of $\delta_{c0}$ for different pairs of
values  $\alpha$ and $\gamma$ with $<N_s>=10$. This is the case
associated to Eq.(\ref{power2}). } \label{ronca}
\end{figure}

\begin{figure}[!h] \centering \resizebox{9cm}{!}
{\rotatebox[origin=c]{0}{\includegraphics{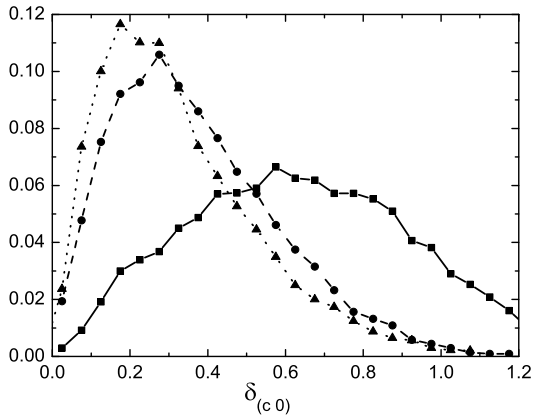}}} \caption{
Distribution of $\delta_{c0}$ for different pairs of values
$\alpha$ and $\gamma$ with $<N_s>=10$. This is the case associated
to Eq.(\ref{power3}). } \label{roncl}
\end{figure}

Figures \ref{ronca} and \ref{roncl} show the distribution of
$\delta_{c0}$ for different values of $\alpha$ and $\gamma$ for
the cases described by Eqs. (\ref{power2}) and (\ref{power3})
respectively. In this case there is an evident difference in the
behaviour of the system. The values of $\alpha$ and $\gamma$ were
adjusted to get a final configuration with the same number of
surviving countries in each case. We can observe that if we
consider local aspect related to the military power and the cost
of defending the borders we get  capitals located closer to the
center of mass of the states.

\begin{figure}[!h] \centering \resizebox{9cm}{!}
{\rotatebox[origin=c]{0}{\includegraphics{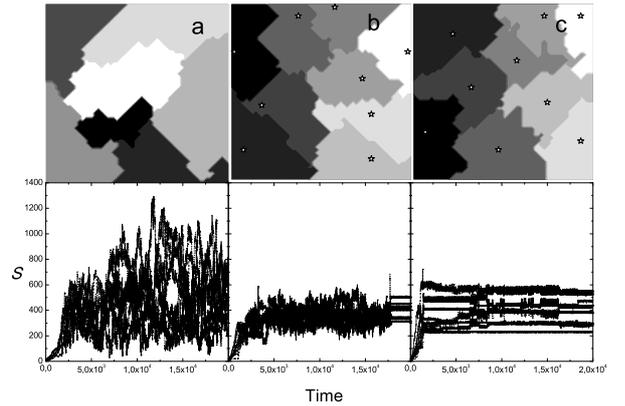}}}
\caption{Maps and size evolution of surviving countries for three
different cases } \label{secuenc}
\end{figure}

To provide a more graphical example of the results we present the
outcome of three single realizations  in Fig \ref{secuenc},
together with a representation of the time evolution of the sizes
of the surviving countries.
Each realization corresponds to a different choice for $P$. The
group (a) corresponds to the model that does not take into account
the location of capitals, Eq. (\ref{power1}), while (b) and (c)
differ in that the last one reflects the state of a system where
$P$ is affected by local contributions, Eq. (\ref{power3})and (b)
corresponds to  Eq. (\ref{power2}). When we observe the temporal
evolution of the size of the surviving countries the differences
are rather evident. The curves related to model associated to Eq.
(\ref{power1}) present fluctuations of great amplitude. This
behaviour reflects in the distribution of sizes of the states than
can be observed in the map showing an instantaneous picture of the
system. At a given moment some countries can be six times bigger
than others. If fluctuations are strong enough they can lead the
system to a trivial state. On the other hand, groups (b) and (c)
show a more steady behaviour. After a transient, the size of the
countries stabilizes suffering only very small fluctuations. At
the same time the distribution of sizes is more even. Groups (b)
and (c) differ en that the effect of local considerations is the
centralization of the capitals, displayed with stars on the map.
This effect was already discussed in Fig.\ref{roncl}.

\section{Conclusions}

Supported by the fact that the historical evolution of geopolitics
is a complex process involving a huge variety of causes and
effect, the conception of simple models explaining  general
aspects has been neglected. This fact is the main motivation
behind the present work. It is not the intention of this model to
provide an accurate and complete description of the process that
shaped the present political division of the world. On the
contrary, its more modest goal is to show that some general
features can be explained departing from very simple assumptions.
Based on the ideas sketched by Collins in (Collins 1995, Collins
1978)  we have developed a model that succeeded to reproduce a
series of geopolitical phenomena, namely the fact that most of the
capitals are located in a central position, the fact that most of
the surviving countries have cost, with only a few being internal,
the possibility of stabilizing a political division with many
countries of similar size and overall shape, but also describing
the situation when the rise of an empire is possible. Still the
model needs some adaptation to be able to describe the sort of
processes that have occurred in The Americas and Africa, and
include the possibility of state segregation. These aspects will
be taken into account ia a future work.

\end{document}